\documentstyle[psfig]{mn}

\newcommand{\eq}{\begin{equation}}
\newcommand{\en}{\end{equation}}

\def\deg{^{\circ}}
\def\etal{{\it et al.}\thinspace}
\def\ie{{\it i.e.,}\thinspace}
\def\eg{{\it e.g.,}\thinspace}
\def\apj{{ApJ}\thinspace}
\def\aap{{A\&A}\thinspace}
\def\mnras{{MNRAS}\thinspace}
\def\nat{{NAT}\thinspace}

\begin{document}

\title[PSR 2303+30]{PSR 2303+30: A single system of drifting subpulses, moding, and nulling.}

\author[Redman, Wright $\&$ Rankin]{
Stephen L. Redman$^{1}$, Geoffrey A.E.Wright$^{1,2}$  \& Joanna M.Rankin$^{1}$\\
$^1$Physics Department, University of Vermont, Burlington, VT 05405 USA : stephen.redman@uvm.edu, joanna.rankin@uvm.edu\\
$^2$Astronomy Centre, University of Sussex, Falmer, BN1 9QJ, UK : gae@sussex.ac.uk}

\date{Received.....Accepted.........; in original form 2003...}

\maketitle

\begin{abstract}
Analyses of multiple pulse sequences of B2303+30 reveal that this pulsar has two distinct emission modes.  One mode (B) follows a steady even-odd pattern and is more intense.  The second mode (Q) is characteristically weak, but has intermittent driftbands with a periodicity of approximately 3$P_{1}$/cycles, and nulls much more frequently than the B mode.  Both modes occur with roughly equal frequency and their profiles have a similar single-humped form with a slight asymmetry. Our observations and analyses strongly suggest that the subpulse driftrates in both modes are linked in a series of cycles, which can be modelled as relaxing oscillations in the underlying circulation rate.   
\end{abstract}


\section{INTRODUCTION}
The phenomena of drifting, moding, and nulling are not new concepts to the pulsar community.  However, rarely has a pulsar offered such insight into its own inner workings through the interaction of these three phenomena.  Not only does PSR B2303+30 provide us with numerous examples of drifting subpulses, moding, and nulling, but it also offers enough clues to model its behavior empirically, as a single system.

PSR B2303+30 is an old, bright pulsar with a relatively long rotation period ($P_{1}$) of 1.57 seconds.  Since its discovery by Lang in 1969, it has been the subject of a number of studies which emphasized measurements of $P_{2}$, the longitudinal distance between adjacent subpulses, and $P_{3}$, the number of rotation periods required before the subpulses return to their original phase (Taylor \& Huguenin 1971; Backer 1973; Sieber \& Oster 1975).  It is one of seven well-known pulsars with a clear even-odd modulation --- of which B0943+10 is also a prominent member --- the others being B0834+06, B1632+24, B1933+16, B2020+28, and B2310+42 (Oster \etal\ 1977; Wolszczan 1980; Rankin 1986; Hankins \& Wolszan 1987).  Careful analyses of B0943+10 have led to the delineation of its rotating subbeam configuration (Deshpande \& Rankin 1999, 2001).  Given the similarities between B0943+10 and B2303+30, it seems reasonable to assume that similar analyses might reveal greater details concerning B2303+30's drift behavior.  However, B2303+30's drift behavior is much more complicated than that of B0943+10.  B2303+30's brightest drifting subpulses do not often have a single steady driftrate, but instead appear to oscillate about $P_{3}\approx{2}$ periods/cycle (hereafter $P_{1}$/cycle or $P_{1}/c$).  Various models have been applied to this strange feature (Gil \etal\ 1992; Gil \& Sendyk 2000) --- however, we are confident that our latest analyses provide sufficient reasons for advancing a more comprehensive explanation.

\begin{figure}
\begin{flushleft}
\begin{tabular}{@{}lr@{}lr@{}}
{\mbox{\psfig{file=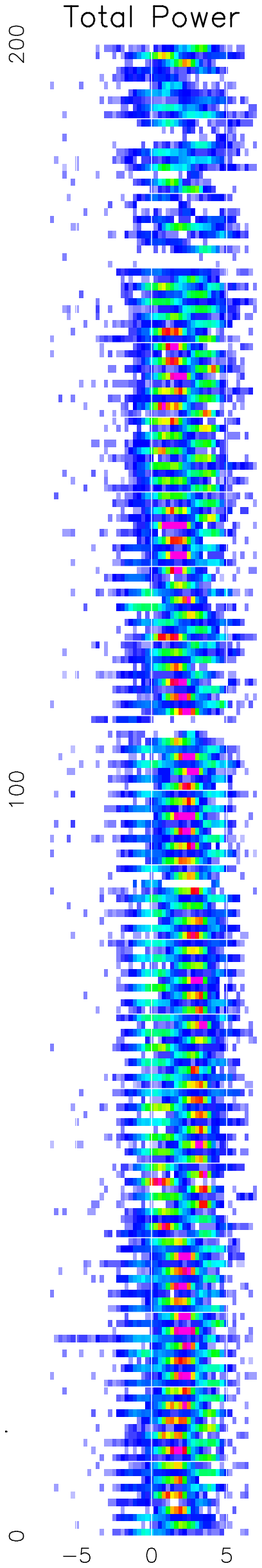,height=19.5cm}}}&
{\mbox{\psfig{file=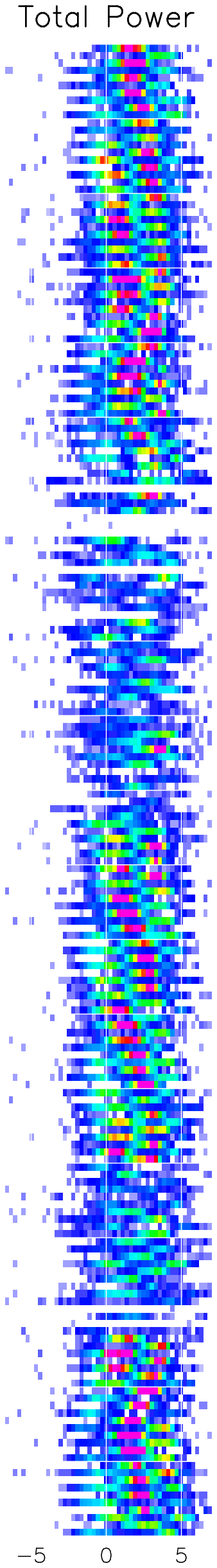,height=19.5cm}}}&
{\mbox{\psfig{file=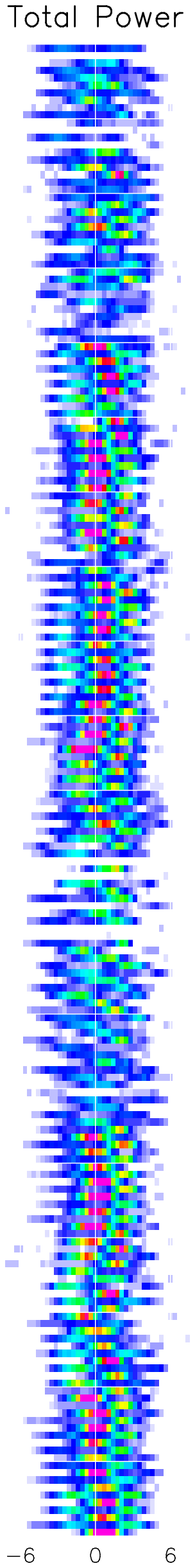,height=19.5cm}}}\\
\end{tabular}
\end{flushleft}
\caption{Three different 200-pulse, total-power sequences of B2303+30, where pulse number is plotted against the longitude in degrees via a colour scale where blue increases to green and red.  The left and center plots are from 430-MHz observations and the plot on the right is the October 7 observation at 327 MHz.  Note the exceptionally long B-mode sequence ($P_{3}\approx{2} P_{1}/c$) in the left panel, the varying $P_{3}$ of the B-mode sequences in the center panel, and the Q-mode intervals ($P_{3}\approx{3} P_{1}/c$) which interrupt the B mode in the right panel.}
\label{Fig1}
\end{figure}

A glance at a typical B2303+30 sequence will confirm without a doubt that it exhibits a strong even-odd modulation ($P_{3}\approx{2}P_{1}/c$) with considerable variation (see Figure~\ref{Fig1} for three sample total-power sequences).  In addition to the clear even-odd modulation, there is a second behavior that, when viewed under good signal-to-noise conditions (hereafter, S/N), sometimes reveals a steady $P_{3}\approx{3}P_{1}/c$.  Because such intervals are characterized by frequent nulls and sometimes erratic behavior, they were difficult to discern in older observations.  This change in behavior is quasi-periodic, suggesting that regular events in the pulsar's magnetosphere or on the crust of the neutron star disturb the drift.  Because these behavior changes are so frequent, B2303+30 presents us with a unique opportunity to use mode changes as a method for modeling the drifting subpulse behavior.  

We therefore argue that there are two principle modes in B2303+30, which can be identified based on their modulation patterns and total-power intensities.  The first, which we call the `B' (burst) mode, has a modulation pattern with a $P_{3}$ very close to 2$P_{1}/c$ and is relatively luminous compared to the second mode, which we call the `Q' (quiescent) mode.  The Q-mode pattern has a $P_{3}$ close to 3$P_{1}/c$, is dimmer than the B mode, and also nulls much more frequently.  These specific modulation patterns are possible only under certain alias conditions, which allow us to posit a basic model of the subbeam rotation.  This model unites the basic features of drift, moding, and nulling as a single cyclic system.

After providing the details of our observational methods in \S 2, we describe our basic method of behavioral analysis in \S 3, in which we suggest that the two behaviors represent two distinct emission modes in B2303+30.  Null analyses of the two modes are covered in \S 4, followed by the deduced geometry of B2303+30 in \S 5 and our drifting subbeam model in \S 6.  In \S 7, we discuss how these implications relate to magnetospheric theories.  Finally, in \S 8, we review our conclusions.

\section{Observations}
The analyses in this paper are based on six single-pulse observations, all of which were carried out at the Arecibo Observatory in Puerto Rico over the past 12 years.  All of these observations measured the four Stokes parameters, $I$, $Q$, $U$, and $V$ to varying degrees of accuracy.  Details on the calibration and polarimetry of the 1992 observations can be found in Hankins \& Rankin (2004).

\begin{center}
\begin{tabular}{|l| |c| |c| |c|}
\hline
\multicolumn{4}{|c|}{Observations}
   \\ \hline\hline
\multicolumn{1}{|c|}{Date}
& \multicolumn{1}{|c|}{Frequency}
& \multicolumn{1}{|c|}{Pulses}
& \multicolumn{1}{|c|}{Bandwidth}
\\ \multicolumn{1}{|c|}{}
& \multicolumn{1}{|c|}{(MHz)}
& \multicolumn{1}{|c|}{}
& \multicolumn{1}{|c|}{(MHz/\#Chans)}
	\\ \hline
1992 October 15 & 430 & 2370 & 10/32  \\ \hline
1992 October 18 & 1414 & 1470 & 20/32 \\ \hline
1992 October 19 & 1414 & 1008 & 20/32 \\ \hline
2003 July 17 & 1525 & 2815 & 100/32 \\ \hline
2003 October 7 & 327 & 1525 & 25/256 \\ \hline
2003 October 20 & 327 & 1523 & 25/256 \\ \hline
\end{tabular}
\end{center}

The 1992 observations have a resolution of 0.275$^{\circ}$ longitude, the first two 2003 observations have a resolution of 0.352$^{\circ}$ longitude, and the 2003 October 20 observation has a resolution of 0.468$^{\circ}$ longitude.

\begin{figure}
\begin{flushleft}
\begin{tabular}{@{}lr@{}lr@{}}
{\mbox{\psfig{file=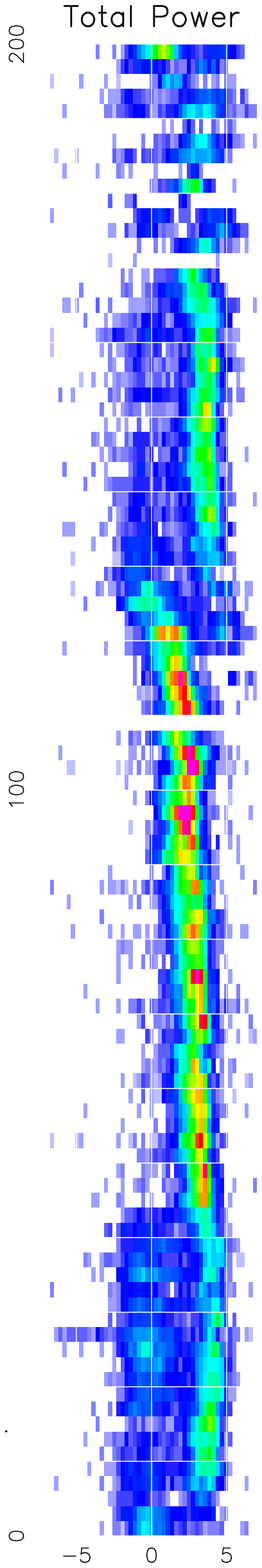,height=19.5cm}}}&
{\mbox{\psfig{file=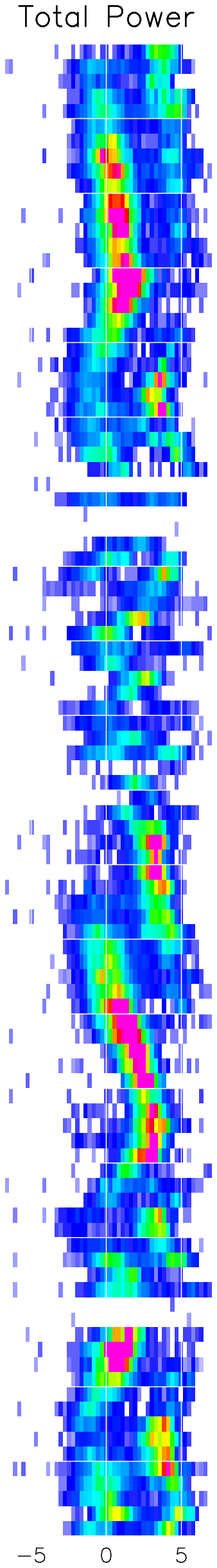,height=19.5cm}}}&
{\mbox{\psfig{file=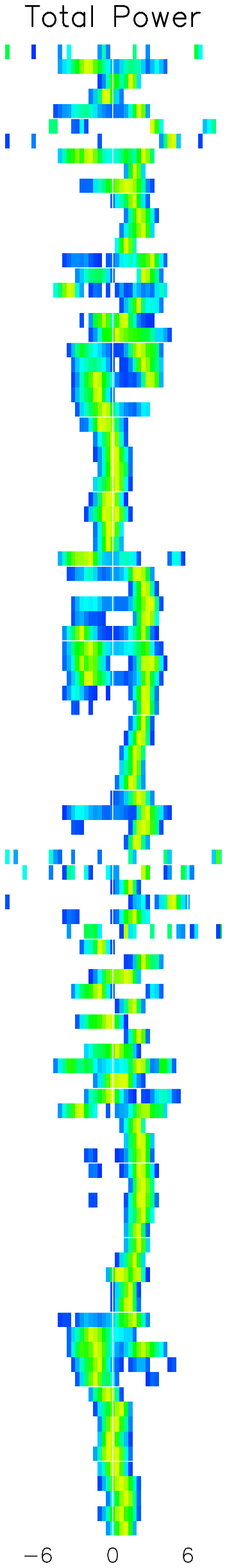,height=19.5cm}}}\\
\end{tabular}
\end{flushleft}
\caption{The same three 200-pulse, total-power sequences seen in Figure~\ref{Fig1}, but with only the odd pulses, plotted at double height.  The plot on the far right, at 327 MHz, is normalized so that all pulse peaks appear the same colour.  In these plots, the B-mode drift appears steady, while the Q mode appears disordered.  Notice that there appear to be at least two different kinds of B-mode behavior: one in which the modulation is exactly 2 $P_{1}/c$ (as between pulses 1-39 in the left panel), and another where the B-mode appears to oscillate around (but not at) 2 $P_{1}/c$ (as between pulses 143-197 of the center panel).}
\label{Fig16}
\end{figure}

\section{The two emission modes}

Figure~\ref{Fig1} shows three different 200-pulse total power sequences.  The left and center plots are from a 430-MHz observation.  The plot on the right is from a 327-MHz observation.  These same three pulse sequences (hereafter PSs) can be seen in Figure~\ref{Fig16}, where only the odd pulses are plotted, at double height.  With these displays, the apparent direction and speed of the drifting subpulses becomes much clearer.  

In the first 430-MHz PS, we see an unusually long B-mode interval (pulses 1 --- 170).  Note the steady even-odd modulation that appears to drift towards the leading edge of the profile, especially near pulse 170.  After two nulls, this even-odd modulation switches to an apparently chaotic behavior with no obvious drift direction.  This is the behavior noted by Rankin \etal\ (1974) and Gil \etal\ (1990).

In the middle sequence (also at 430 MHz), the B mode is much less steady, changing apparent drift direction frequently.  The short drift towards the right (pulses 33-41) has a $P_{3}$ with the characteristic B value of $\approx{2}P_{1}/c$, but is relatively weak.  There are also two examples of what appears to be an `oscillation' of $P_{3}$ (between pulses 33-95 and 143-197).  It is this slight variation in $P_{3}$ which Gil \etal\ (1990) identified as a second B mode.  This variation in $P_{3}$ often occurs immediately after a transition to the B mode.  Between the two B-mode sequences, another chaotic PS appears.

\begin{figure}
$$\vbox{
\psfig{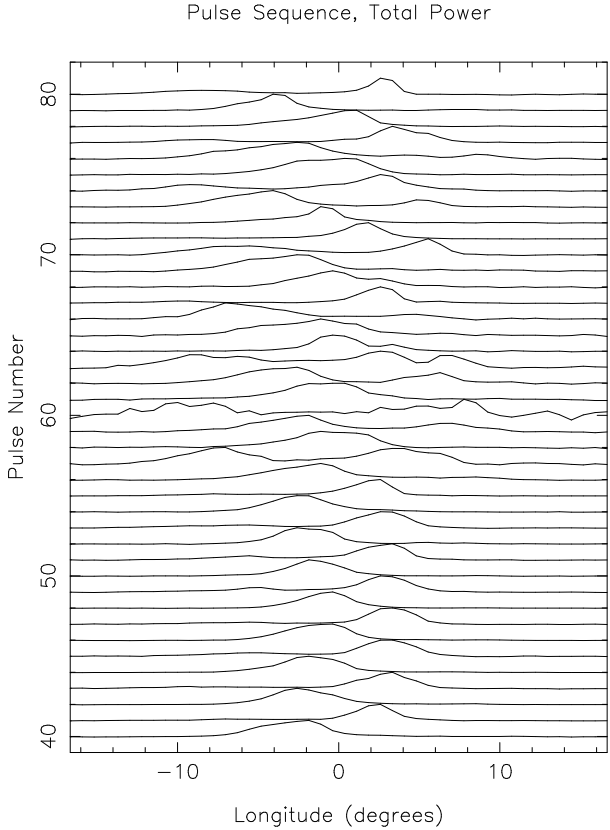}
}$$
\caption{Pulses 40 through 80 of the 327-MHz observation (also seen in the right-hand panel of Fig.~\ref{Fig1}), normalized so that all pulses peak at the same height.  The two modes can be identified by their respective driftrates.  Pulses through number 55 or 56 are B mode, and the Q mode finishes the sequence.  Pulse 60 is not a null, but a weak pulse.}
\label{Fig2}
\end{figure}

In the 327-MHz sequence, the larger S/N makes the drift behavior of the seemingly chaotic intervals much more apparent than in former observations.  The B mode begins the sequence, then switches to a steady $P_{3}\approx{3}P_{1}/c$ PS around pulse 55.  This transition can be seen in greater detail in Figure~\ref{Fig2}.  Notice that the transition between behaviors is smooth and unbroken.  This second behavior continues through several nulls until pulse 92, where the pulsar returns to B mode, only to switch back after pulse 160.

It is our conclusion that these ``chaotic'' sequences represent a second mode, the `Q' mode.  While this second behavior appeared completely disordered in earlier data, more recent observations reveal that it frequently has a quasi-steady $P_{3}\approx{3}P_{1}/c$.  However, the Q mode also exhibits numerous nulls and can otherwise appear disordered.

A further point should be noted here --- nulls frequently occur at the transition boundaries between modes; although, transitions from the B to the Q mode are also sometimes surprisingly smooth (Fig.~\ref{Fig2}).  This distinction will become significant later when we attempt to model the drift behavior in \S 6.

\subsection{Mode Identification and Statistics}

After displaying and studying each PS, we attempted to identify the mode of each pulse.  PSs were categorized as either B- or Q-mode pulses based upon the $P_{3}$ of the sequence.  This procedure was carried out entirely by eye, using coloured displays such as those seen in Figs.~\ref{Fig1} \&~\ref{Fig16}.

This could not, admittedly, be accomplished without some subjectivity.  For example, in Figure~\ref{Fig2}  a smooth transition from the B to the Q mode can be seen at pulse 56.  Do we classify pulse 56 as a B- or a Q-mode pulse?  Or, if a null occurs between modes, as it often does, do we classify the null as belonging to the B or the Q mode?  Clearly, for some analyses, such as constructing partial modal profiles, a decision such as this would make little difference.  However, the issue becomes more complicated if, for example, we classify all transition nulls as belonging to the Q mode and then later conclude that the Q mode nulls more frequently than the B mode.  For the purpose of such accounting then, the transition pulses (or nulls) were simply ignored.

The occurrence frequency of the two modes is consistent among the available observations.  45.7 and 54.3$\pm{1}$\% of the 9,188 pulses we observed are Q-mode and B-mode pulses, respectively.  The average length of a B-mode sequence is 37 pulses, and that of a Q-mode 31 pulses.

\begin{figure}
$$\vbox{
\psfig{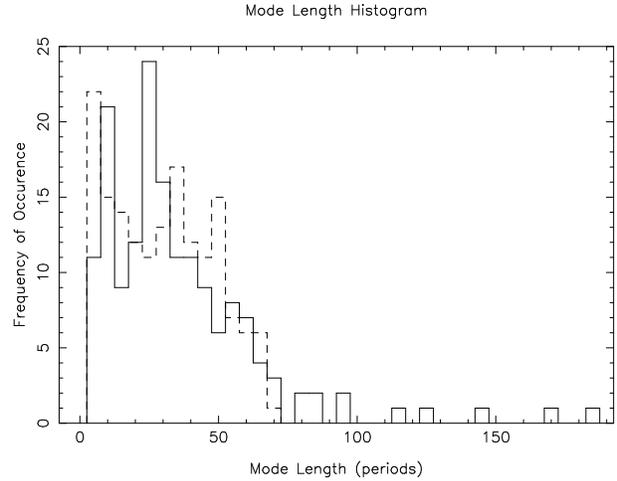}
}$$
\caption{Mode lengths, from all observations --- B mode, solid; Q mode, dashed.  Notice that the B mode can last for over a hundred periods, but no Q-mode PS longer than 71 periods was observed.  Also note that for both modes, the distribution is bi-modal.}
\label{Fig24}
\end{figure}

Figure~\ref{Fig24} shows a histogram of B- and Q-mode lengths.  Notice that the B mode sometimes persists for very long stretches, some of which are well over 100 periods --- but the Q mode never seems to last for more than about 70 pulses.  Figure~\ref{Fig24} also suggests that the mode lengths follow a bimodal distribution --- that is, a population of both short and long PSs for each mode.  The short B-mode sequences are crucial for understanding the subbeam circulation model presented in \S 7.

\subsection{Integrated profiles}

\begin{figure}
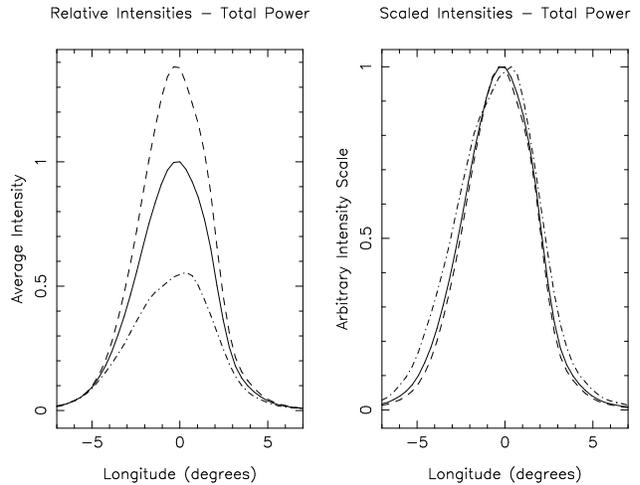

\begin{flushleft}
\begin{tabular}{@{}lr@{}}
{\mbox{\psfig{file=relative_327.ps,width=3.9truecm,angle=-90.}}}&
{\mbox{\psfig{file=scaled_327.ps,width=3.9truecm,angle=-90.}}}\\
\end{tabular}
\caption{B-mode (dashed), total (solid), and Q-mode (dash-dotted) total-power profiles at 327 MHz (October 7 observation).  The average-profile peak of the Q mode is about 43\% that of the B-mode between 0.3 and 1.6 GHz.  The figure on the right shows the same profiles, all scaled to the same arbitrary height.  Notice that the Q-mode profile is noticeably wider than its B-mode counterpart.}
\label{Fig3}
\end{flushleft}
\end{figure}

\begin{figure}
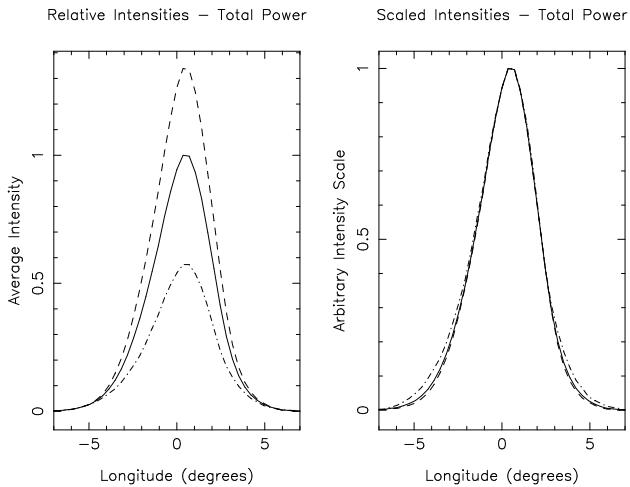

\begin{flushleft}
\begin{tabular}{@{}lr@{}}
{\mbox{\psfig{file=relative_1575.ps,width=3.9truecm,angle=-90.}}}&
{\mbox{\psfig{file=scaled_1575.ps,width=3.9truecm,angle=-90.}}}\\
\end{tabular}
\caption{Total-power profiles as in Fig.~\ref{Fig3} at 1525 MHz.  Again, note that the Q-mode is both weaker and slightly wider than the B-mode profile.}
\label{Fig4}
\end{flushleft}
\end{figure}

Partial total-power profiles, comprised of pulses identified as belonging to the B and Q modes, were first computed so that their relative average intensity might be compared, and then scaled so that the profiles had the same height.  The scaled profiles allowed us to compare the shapes of each modal partial profile.  Figs.~\ref{Fig3} \&~\ref{Fig4}, at 430 and 1525 MHz, respectively, each show a relative and scaled profile of B2303+30.

In both figures, two circumstances can be easily confirmed.  First, the B mode is significantly brighter than the Q mode.  Among the six observations, the B mode was, on average, 2.32$\pm{0.31}$ times as intense as the Q mode.  Second, the Q-mode profile is noticeably wider than its B-mode counterpart.  The causes of this half-width increase are, at this point, not well understood.  However, it can be seen that this increase is less pronounced at the higher frequency.

It should also be noted that B2303+30 has a slightly asymmetric average profile --- the trailing edge is slightly steeper than the leading edge of the profile.

Information about the polarization-modal structure of B2303+30's profiles can be found in Ramachandran \etal\ (2002), as well as Rankin \& Ramachandran (2003) and Hankins \& Rankin (2004).

\subsection {Intensity Variations}

\begin{figure}
$$\vbox{
\psfig{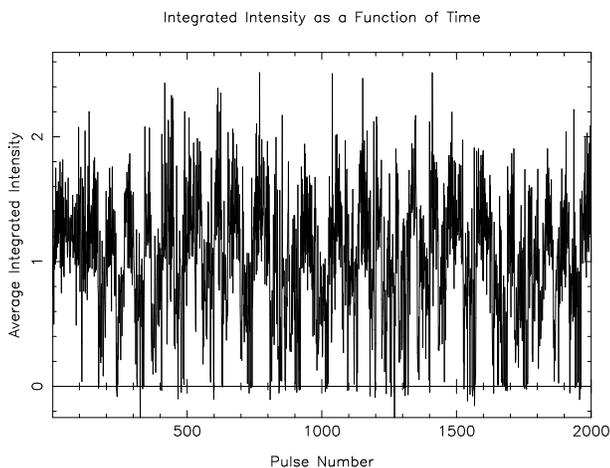}
}$$
\caption{The average integrated intensity of a sequence of single pulses at 430 MHz.  Note the pronounced intensity `states' of the brighter B and weaker Q modes --- in roughly a ratio of 2 to 1.  One can even clearly pick out the exceptionally long B-mode PS which begins this observation.  Note also how the B and Q modes alternate with a quasi-regular periodicity.}
\label{Fig6}
\end{figure}

\begin{figure}
$$\vbox{
\psfig{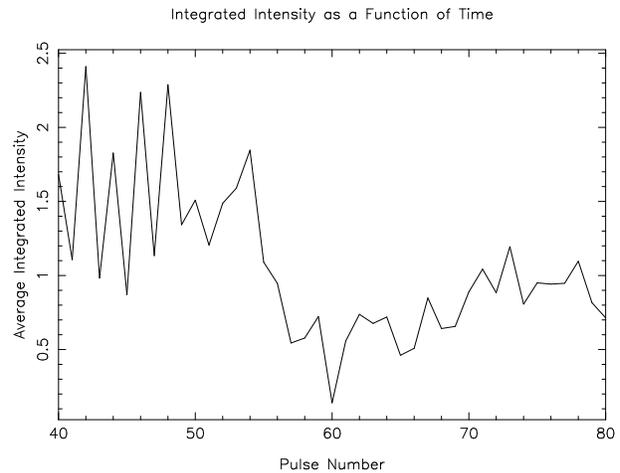}
}$$
\caption{Intensities of the same 40-pulse interval seen in Fig.~\ref{Fig2}.  Here the two modes can be identified by their respective total intensities.  Notice the clear drop in intensity around pulse 55, coinciding with the change in driftrate seen in Fig.~\ref{Fig2}.}
\label{Fig7}
\end{figure}

Having discovered how dramatically different the average intensities of the modes are, we now examine the long-term intensity variations.  Figure~\ref{Fig6} gives the relative integrated intensity of a 2000-pulse sequence at 430 MHz.  The clear and quasi-periodic variations in intensity are not the result of scintillation; it is a feature of B2303+30 that is seen at all observed frequencies.  As might be expected from the previous subsection, these variations in integrated intensity result from the fact that the pulsar has two modes, each with a characteristic intensity.

The mode-dependent intensity variations are even more clearly demonstrated when normalized PSs are compared with the integrated intensity of the same sequence.  Figure~\ref{Fig7} shows the same 40-pulse sequence seen in Figure~\ref{Fig2}.  Notice that there is a pronounced intensity decrease around pulse 55 in Figure~\ref{Fig7}, corresponding exactly to the change in $P_{3}$ shown in Figure~\ref{Fig2}.  Thus, these changes in intensity are characteristic of each mode.  

These changes in intensity also represent a second method of identifying the modes.  This method is less reliable than distinguishing between the modes using $P_{3}$, however, and therefore was only used to clarify a possible mode change.


\subsection{Subpulse modulation}

\begin{figure}
$$\vbox{
\psfig{file=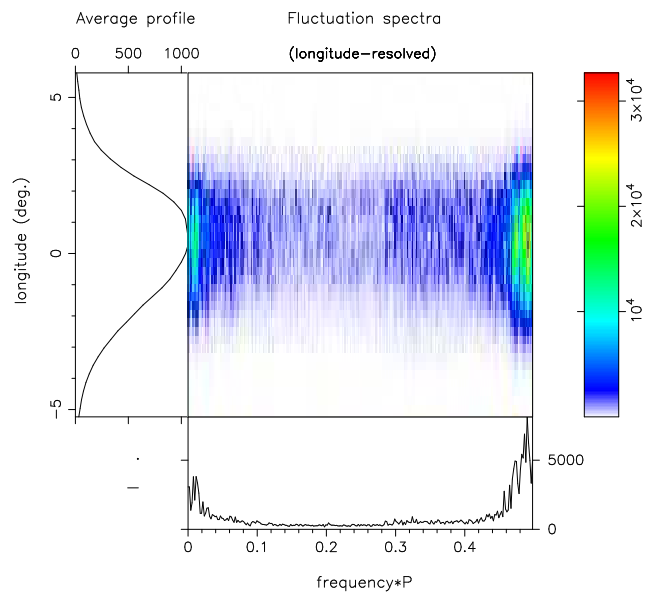,width=8truecm}
}$$
\caption{Longitude-resolved fluctuation spectra for pulsar 2303+30 at 430 MHz.  A 512-point FFT was used and averaged over all 2370 pulses of the 1992 October 15 observation.  The body of the figure gives the amplitude of the features, according to the colour scheme on the right.  The average profile (Stokes parameter $I$) is plotted in the left-hand panel, and the integral spectrum is given at the bottom of the figure.  Note the numerous features, particularly those around 0.005, between 0.29-0.45, and 0.45-0.49 c/$P_{1}$.}
\label{Fig9}
\end{figure}

\begin{figure}
$$\vbox{
\psfig{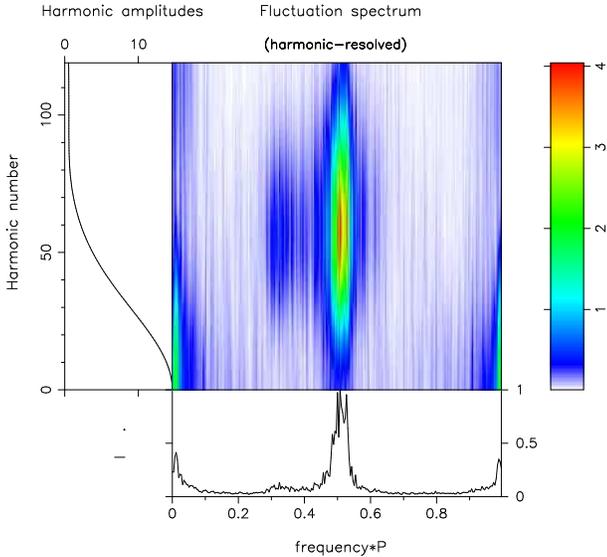}
}$$
\caption{A harmonically-resolved fluctation spectrum of all 2370 pulses in the 430 MHz observation from 1992 October 15.  The lefthand panel gives the amplitude of frequency components at integral multiples of the pulsar rotation frequency, 1/$P_{1}$.  The body of the figure then gives the amplitude of all the other frequency components in the spectrum up to 120/$P_{1}$ according to the colour scheme at the right.  The bottom panel shows the sum of these frequency components, collapsed onto a 1/$P_{1}$ interval.  The features now fall at their true (partially unaliased) frequency.  The B-mode features can be seen around 0.5 c/$P_{1}$, corresponding to a $P_{3}\approx{2} P_{1}/c$.  The Q-mode feature, with a $P_{3}\approx{3}P_{1}/c$, can be seen to the left of the B-mode feature.  There are also two other features to be considered in this diagram --- one close to 0.01 c/$P_{1}$, and a second near 0.99 c/$P_{1}$ --- which represent an amplitude modulation that occurs every 80 pulses or so, the approximate sum of a given B and Q mode pair.  This amplitude modulation is apparent in Fig.~\ref{Fig6}.}
\label{Fig8}
\end{figure}

\begin{figure}
$$\vbox{
\psfig{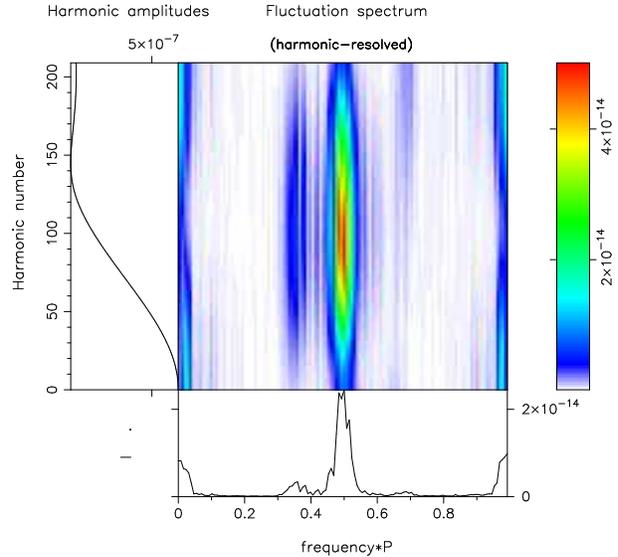}
}$$
\caption{A harmonically-resolved fluctuation spectrum of 1000 simulated pulses.  Notice the similarity of features seen in Fig.~\ref{Fig8}, such as the strong B-mode feature, the weaker Q-mode feature, and a strong amplitude modulation at either end of the frequency spectrum.}
\label{Fig80}
\end{figure}

As indicated at the beginning of \S 3, the two modes each exhibit a unique but varying $P_{3}$.  The longitude-resolved fluctuation spectra (hereafter, LRF spectra) of all 2370 pulses at 430 MHz (see Figure~\ref{Fig9}) reveals the complexity of B2303+30's subpulse modulation.  Strong features are visible in several locations throughout the integral spectrum (bottom panel).  The origin of these features will be discussed below.


From the LRF spectra, it is impossible to determine whether the features are indicative of the pulsar's true frequencies, or aliases thereof.  The aliasing question can sometimes be resolved however, by computing a harmonically-resolved fluctuation spectrum (hereafter, HRF spectrum).  The HRF spectrum is calculated by overlapping fast Fourier transforms (FFTs) of length 256 and then interpolating between adjacent Fourier amplitudes to estimate the frequency of the feature and its errors [For further details see Deshpande \& Rankin (2001)].  HRF spectra can further sometimes distinguish between amplitude and phase modulations.  Amplitude modulations appear as symmetric feature pairs in the integral spectrum, whereas phase modulations are asymmetric.  The beautiful HRF spectrum of the entire 430-MHz sequence, computed with an FFT of 256 points, can be seen in Figure~\ref{Fig8}.

To determine the sources of these features, a simple pulsar model was computed, complete with strong B-mode ($P_{3}\approx{2}P_{1}/c$) and weak Q-mode ($P_{3}\approx{-3}P_{1}/c$) emission, each with moderate variation in their $P_{3}$ values.  Mode lengths were also variable throughout the sequence, lasting between 11 and 23 periods.  Neither nulls nor noise were included in the simulation.  The resulting HRF of the simulation can be see in Figure~\ref{Fig80}.  Notice that the four features mentioned above appear here in decent detail.

By altering and eliminating certain variables (such as the length of the modes, or the intensity of the emission), we could discern the source of those features seen in Fig.~\ref{Fig8}.  

The first is that seen near 0.5 c/$P_{1}$.  This feature is due primarily to the B-mode modulation.  Notice, though, that this feature is not smooth, but comprised of high-$Q$ (=$f/\Delta f$) `spikes' --- produced by both the variety of semi-discrete $P_{3}$ values the B-mode exhibits, as well as the fact that there are two prominent $P_{3}$ values in this pulsar, the combination of which produces periodic `beats' in the frequency spectrum.

The second of the four prominent features seen in Figure~\ref{Fig8} can be seen from 0.28-0.44 c/$P_{1}$.  This feature, which is clearly visible in the body of the figure, is produced by Q-mode emission, as well as the above-mentioned `beat' phenomenon.  Notice that this feature is not reflected about 0.50 c/$P_{1}$, and thus represents phase modulation.  The simulation verifies that the location of this feature indicates a specific drift direction (\ie negative).

The third and fourth features in Figure~\ref{Fig8} are reflections of each other --- one at 0.012 c/$P_{1}$, and the other at 0.998 c/$P_{1}$ --- corresponding to an amplitude modulation of about $89.08\pm{1.24} P_{1}/c$.  These features are due to the amplitude modulation that occurs between successive B- and Q-mode PSs (\eg Fig.~\ref{Fig6}).

\section{Nulls}

As can be seen in the three PSs of Fig.~\ref{Fig1}, B2303+30 occasionally emits at such low intensity that the pulsar appears to null.  These low-intensity intervals appear to be much more frequent during Q-mode intervals, but a few can be found within B-mode sequences as well.  Transition intervals of low intensity between modes also occur throughout the PSs, as can be seen, for example, in the left-most panel of Fig.~\ref{Fig1}.

\subsection{Nulls or Weak Pulses?}

\begin{figure}
$$\vbox{
\psfig{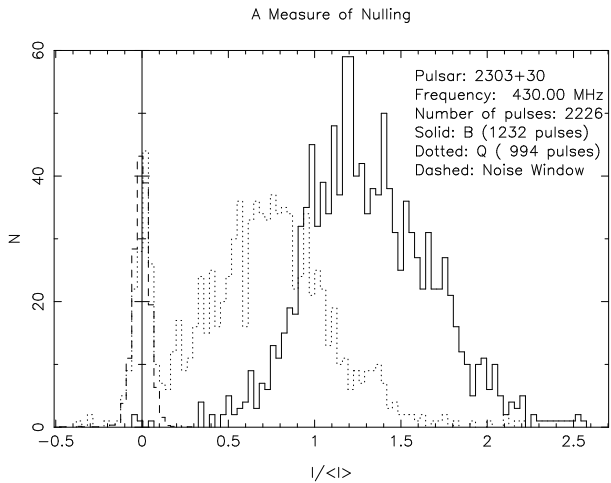}
}$$
\caption{Null-frequency histogram for the B- and Q-modes.  Transition pulses have been ignored for reasons discussed in \S 3.1.  Notice that the small population of B-mode nulls is well-separated from the B-mode pulses, but that the Q-mode pulse and null distributions overlap slightly.  Also note that the Q-mode nulls have nearly the same distribution as the off-pulse noise (which has been scaled down by a factor of 15 --- the ratio of pulses to Q-mode nulls --- to match the height of the Q-mode null distribution).}
\label{Fig12}
\end{figure}

To determine whether these low intensity intervals are nulls or simply low-intensity pulses, we computed a histogram of the average-pulse intensity.  Figure~\ref{Fig12} shows this histogram, segregated by modes.  For reasons discussed in the introduction, 139 transition pulses have been excluded from this analysis.  Pulses that occur at or very near zero average integrated intensity --- that is, that are indistinguishable from the noise distribution --- are considered nulls.  From this figure, we can see that the B mode has well separated populations of nulls and pulses.  The Q mode also shows two distinct populations of nulls and pulses, but there is a slight overlap between the two near 0.15 times the average intensity.  This strongly suggests that B2303+30 does indeed null.

Figure~\ref{Fig12} provides us with a definition for null pulses.  Those pulses with an average on-pulse integrated intensity below 12\% of the average integrated intensity were considered nulls.  This integrated intensity minimum occurred in all of the observations, and thus provided a consistent definition of a null.

To confirm that these pulses were actually nulls, we created average profiles of null pulses (including transition nulls) at different frequencies.  These profiles (not pictured) revealed that there was no significant   power in those pulses that were considered nulls.

\subsection{Null Statistics}

\begin{center}
\begin{tabular}{|l| |c| |c| |c|}
\hline
\multicolumn{4}{|c|}{Null Statistics}
   \\ \hline\hline
\multicolumn{1}{|c|}{Pulse}
& \multicolumn{1}{|c|}{Frequency}
& \multicolumn{1}{|c|}{Maximum}
& \multicolumn{1}{|c|}{Average}
\\ \multicolumn{1}{|c|}{Population}
& \multicolumn{1}{|c|}{of Occurrence}
& \multicolumn{1}{|c|}{(Periods)}
& \multicolumn{1}{|c|}{(Periods)}
	\\ \hline
All Pulses & 10.02$\pm{2.92}$\% & 11 & 1.84$\pm{0.28}$ \\ \hline
B-Mode & 0.41$\pm{0.41}$\% & 2 & 1.14$\pm{0.19}$\\ \hline
Q-Mode & 18.62$\pm{4.12}$\% & 8 & 1.84$\pm{0.32}$\\ \hline
Transition & 18.22$\pm{9.91}$\% & 11 & 1.51$\pm{0.51}$\\ \hline
\end{tabular}
\end{center}

The above table confirms what Fig.~\ref{Fig12} suggested --- that the Q mode nulls much more frequently than the B mode.  This is the first pulsar to exhibit modes with surprisingly different null fractions.

Transition nulls are nulls that occur between modes, and thus could not be categorized beforehand as belonging to the B or Q modes.  The low frequency of B-mode nulls and the relatively large frequency of Q-mode nulls suggests that those nulls which belong to transition pulses are in fact Q-mode nulls.  This conclusion is further re-enforced by the fact that the null distributions of the Q mode and of the transition pulses overlap.

\begin{figure}
$$\vbox{
\psfig{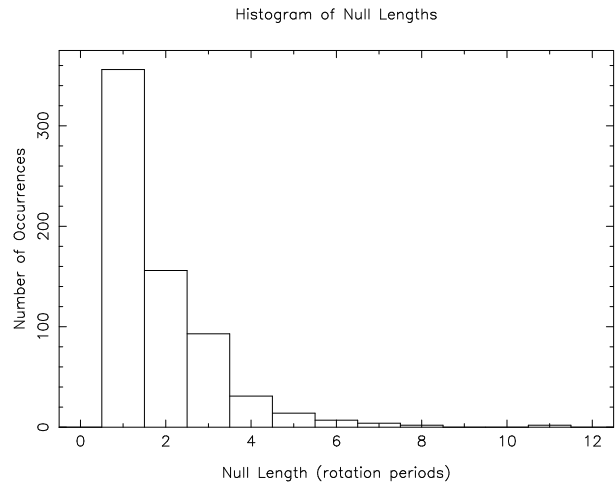}
}$$
\caption{A histogram of null lengths from all available PSs.  Notice that longer nulls are less frequent, indicating that very short nulls (nulls less than one period) probably exist.}
\label{Fig20}
\end{figure}

Figure~\ref{Fig20} shows a histogram of null lengths.  Longer nulls are relatively infrequent compared with short nulls.  This suggests that very short nulls --- nulls less than one period --- probably occur.



\subsection{Subpulse Memory}

When subpulse drift appears to be continuous across a null, the pulsar is said to exhibit `memory'.  The existence of this quality is established by examining the phase of individual pulses before and after a sequence of nulls.  If the pulses have a constant phase change of about $0\deg$ across any number of nulls, the pulsar does not have memory.  Likewise, if the phase change of the pulses shows a constant increase or decrease, drifting continues across nulls, and the pulsar is said to have memory.

\begin{figure}
$$\vbox{
\psfig{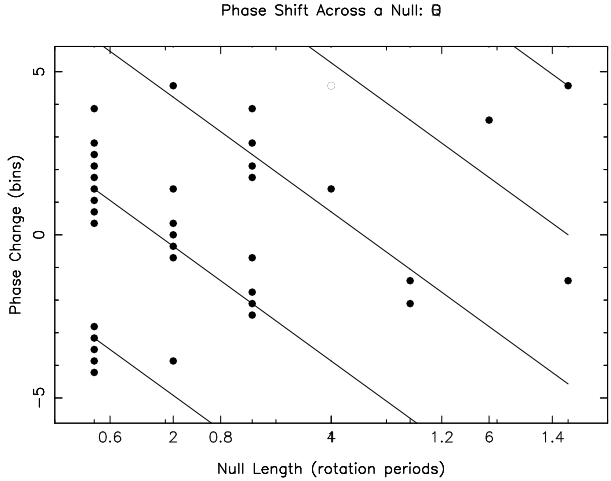}
}$$
\caption{Phase Change across Q-mode nulls at 327 MHz (2003 October 7).  The solid lines represent constant drift across a null.  Since the phase change across a null appears to be dependent upon the length of that null, B2303+30 has memory.}
\label{Fig21}
\end{figure}

Figure~\ref{Fig21} shows a plot of phase change versus Q-mode null length at 327 MHz (2003 October 7 observation).  B-mode nulls have been excluded because they exhibit a different driftrate, and transition nulls have been excluded because they have no consistent driftrate.  Notice that the points appear to follow lines of constant drift, but that they steadily diverge from this drift as the null increase in length.  This indicates that, while the subpulses exhibit very good evidence of having memory, their rotation rates do change speed during nulls.

\section{Geometry and ``Absorption''}

\begin{figure}
$$\mbox{
\psfig{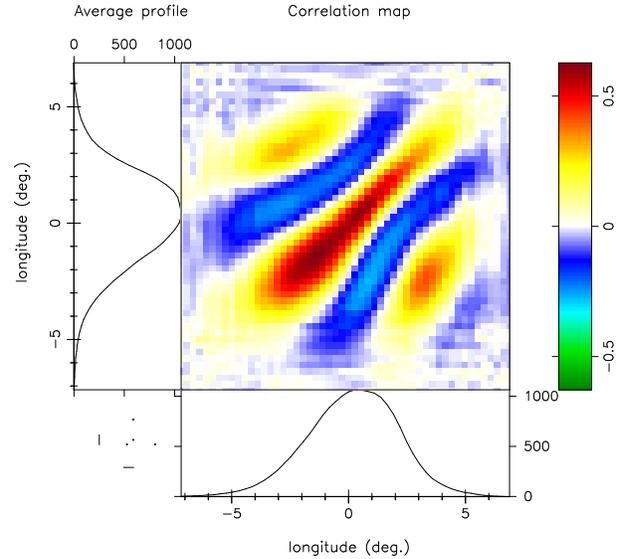}
}$$
\caption{Beautiful evidence that the metre-wavelength profile of 
B2303+30 is not complete.  The figure gives a colour-intensity-coded 
longitude-longitude correlation plot at a lag of two.  Notice that the 
correlation is not symmetrical about the peak of this 430-MHz 
PS's profile (plotted in both the lefthand and bottom panels).  The 
symmetry point rather falls near its trailing half-power point.  The 
vertical or horizontal interval between the maxima is also a measure 
of the subpulse separation $P_2$.  See text.} 
\label{Fig50}
\end{figure}

A crucial question in the analysis of any pulsar exhibiting subpulse drift is its emission geometry --- that is, its magnetic inclination angle $\alpha$ and sightline impact angle $\beta$.  An early attempt to determine these values was made by one of us (Rankin 1993) using a polarization-angle (hereafter PA) sweep rate of 4.5 $\deg/\deg$ (Lyne \& Manchester 1988), resulting in $\alpha$ and $\beta$ values of 20.5 and 4.5$\deg$, respectively.  Unfortunately, we now have confirmed from several directions that such PA sweep rate determinations are particularly difficult for the conal single (${\bf S}_d$) stars most closely associated with subpulse drifting, because of polarization-mode mixing on the edges of the conal beam (Ramachandran \etal\ 2002; Rankin \& Ramachandran 2003; Ramachandran \etal\ 2004)---and, indeed, the PA histogram in the 2002 paper indicates a sweep rate near 8.5 $\deg/\deg$ which has now been confirmed by recent AO observations in this paper.   Keeping the assumption of an outer cone, $\alpha$ would then be some 40$\deg$ without much changing $\beta$.

However, this is not the only issue: (a) Not only do we lack any strong evidence (low frequency profile bifurcation) that the pulsar has an outer cone, the near invariance of its profile (Hankins \& Rankin 2004) might suggest an inner one, and (b) every well studied ${\bf S}_d$ star so far has been found to exhibit some profile ``absorption'', so we cannot be sure that the single profile we observe represents a full traverse through its emission cone.  Indeed, the asymmetry of its profile both in total power and modal polarization also suggest that B2303+30's profile is incomplete on the trailing side, and the longitude-longitude correlation plot in Figure~\ref{Fig50} comes close to proving this circumstance.  Nonetheless, were its full profile even twice as wide (9--10$\deg$) --- corresponding to the magnetic-axis longitude falling at about the half-power point on the trailing profile edge --- $\alpha$ could not be as small as 20$\deg$.

Difficult observations at 100 MHz and below are needed to resolve B2303+30's geometry.  102-MHz observations (Malofeev \etal\ 1989; Suleymanova 2004) suggest an unresolved double form, and certain 111.5 MHz AO profiles from the same era confirm this suggestion (Hankins \& Rankin 2004), but no observation of adequate quality apparently exists to fully resolve whether the star's profile bifurcates in the expected outer cone manner.

None the less, a reasonable guess on the available evidence is that the star has an inner cone and a nearly constant profile half-power width of some 9.5$\deg$ such that only the leading part is seen at metre wavelengths and higher.  The second ``component'' at 100 MHz then corresponds to that ``absorbed'' at higher frequencies.  On this basis, $\alpha$ and $\beta$ are estimated to be about 26$\deg$ and -3$\deg$, respectively.

Finally, we can tentatively conclude that B2303+30 has an ``inside'' or poleward sightline traverse,
or, what is the same, a {\it negative} value of $\beta$.  We follow the argument in Deshpande \& 
Rankin (2001) to the effect that if the PA rotation over $P_2$ (the interval between subpulses) 
$\chi_{P_2} < P_2$, then an inside traverse is indicated.  

\section{Implications and analysis}
Despite attempts to define a basic pulsar emission theory (Ruderman $\&$ Sutherland 1975; Gil \& Sendyk 2000; Hibschmann $\&$ Arons 2001; Harding $\&$ Muslimov 2002; Wright 2003) there remains very little agreement even in explaining drifting subpulses, an elementary observed feature of so many pulsars.  Additional common emission features which frequently accompany subpulse drift such as mode-changing and nulling have received only scant theoretical attention (Jones 1982; Filippenko $\&$ Radhakrishnan 1982; Rankin $\&$ Wright 2003). Thus, in guiding future theory it is up to observers to establish links and correlations between these phenomena.

Here we take a `holistic' view of the behaviour of the radio emission of B2303+30.  Our hope is that by considering the observed phenomena of subpulse drift, moding and nulling of this pulsar as aspects of a $\it{single}$ system we may learn as much from their interactions as from their separate behaviour.

\subsection{The mode system}
Let us briefly recap the principal results needed for our synthesis. We have demonstrated in this paper that the emission system of B2303+30 is dominated by two distinctive modes and their interactions. The modes exhibit rapid transitions from one to another and are usually clearly distinguishable both by their strongly differing intensities and by their altered subpulse behaviour.  We are fortunate that the proportion with which the two modes occur is near to even (54:46 for B : Q) and that their mean durations ($37 P_1$ : $31P_1$) are sufficiently short to give us satisfactory statistics within the observing time available.  However the distribution of  mode durations (see Fig. 4) does not take the simple gaussian form which one might have reasonably expected if there were a random switching between the modes. Firstly, there are occasional stretches of B mode with long and stable duration well over 100 pulses,  which do not belong to the overall pattern. Without these special sequences, the B to Q ratio comes even closer to 50:50, suggesting that for some reason the two modes are, for much of the time, in near-equilibrium. Secondly, the excess of mode durations less than 13 pulses creates a clearly bi-modal distribution. Below we suggest a way in which this bimodality may arise.

Nulls strongly interact with the modes system, since we find in this pulsar --- for the first time among pulsars --- that they overwhelmingly occur in only one of the modes. Nulling is conventionally considered to be a different physical phenomenon to moding. Furthermore, the statistics of null lengths (Fig. 13) are quite different to those of mode lengths (Fig. 4), so that in this and all other pulsars where they occur, short nulls are more frequent and long nulls rare. This contrasts to the peak of 35-40 pulses in the  distribution of B- and Q-mode lengths. However if we only compare the  peak of short-duration modes with the null peaks, then the statistics look far less different. With modes, even more so than with nulls, it is difficult  to identify those of short duration. Indeed it is intrinsically impossible to define anything less than a 3-pulse mode! Taking these factors into account the `true' distribution of the $\emph{short}$ modes may well have the same form as that of nulls, and hence may be part of the same system. Again this is an important clue for our understanding of the system.  

Since the B and Q modes have similar durations, and since these are considerably shorter than the length of the observing sessions, we have the opportunity to test whether or not the system is switching modes at random. If the system is not random, then we might expect correlations between the lengths of successive mode appearances: the length of a B mode sequence may influence the length of the following Q mode (or vice versa), or one B may fix the length of the next B, or the combination of B and Q may form a non-random sequence etc. To test whether some unknown physical `rule' underlies the selection of the modes and the duration of their appearance, we apply a well-known test, developed in the context of chaos theory, to search for order in time-series (Takens 1981).  It involves using the original time-series to create a duplicate time-series with a delay of one unit, and generating a sequence of pairs displayed on a 2-dimensional graph. If the points on the graph show a tendency to be confined to a particular region, or to follow a particular trajectory, then this is strong evidence that the system has a hidden attractor determined by the physical rule. Using the mode lengths as units, it was possible to experiment with several possible kinds of time-series (B-length vs next Q-length, B-length vs next B-length etc.). The typical result was a quasi-cyclic clockwise progression about the mean mode-length, but with sufficient counter-clockwise components to give doubt as to whether the result was due to chance. Often the picture was complicated by the numerous short  mode lengths which contribute to the unexpected peak of these in the histogram of Fig. 4. 

However in one observation at 1414-MHz we found a more convincing result : using the length of successive Q modes a clear cyclic behaviour was revealed (Fig. 16). The S/N of this observation was not as good as in the others and short weak B-mode pulse sequences, which in better observations punctuate the Q mode, were not detected. The histogram of mode lengths for this data set exhibited $\emph{no}$ short modes. Thus the short mode sequences could somehow be clouding the underlying picture of B2303+30's system and should be merely seen as elements of a longer-scale quasi-cyclic behaviour. 

\begin{figure}
$$\mbox{
\psfig{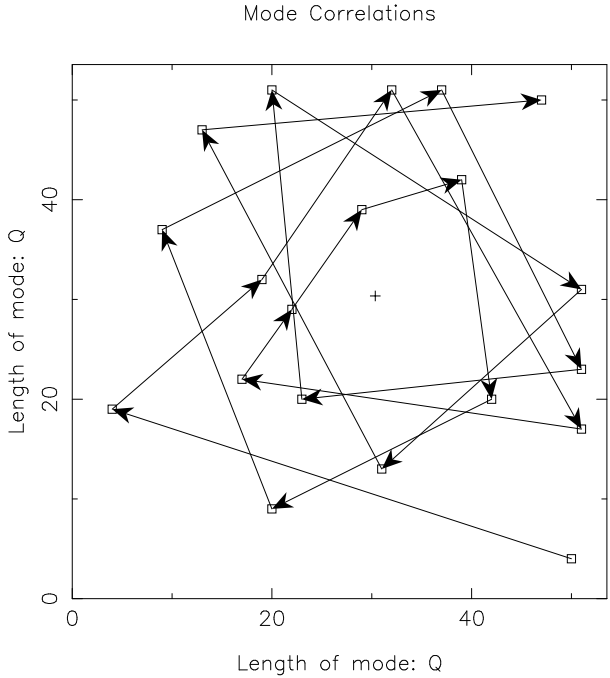}
}$$
\caption{A sequence formed from plotting the length of each Q mode against that of the subsequent Q mode from the 1992 October 18 observation at 1414-MHz. Here a clear quasi-periodic cycle is evident.}
\label{Fig15}
\end{figure}

\begin{figure}
\begin{flushleft}
\begin{tabular}{@{}lr@{}lr@{}}
{\mbox{\psfig{file=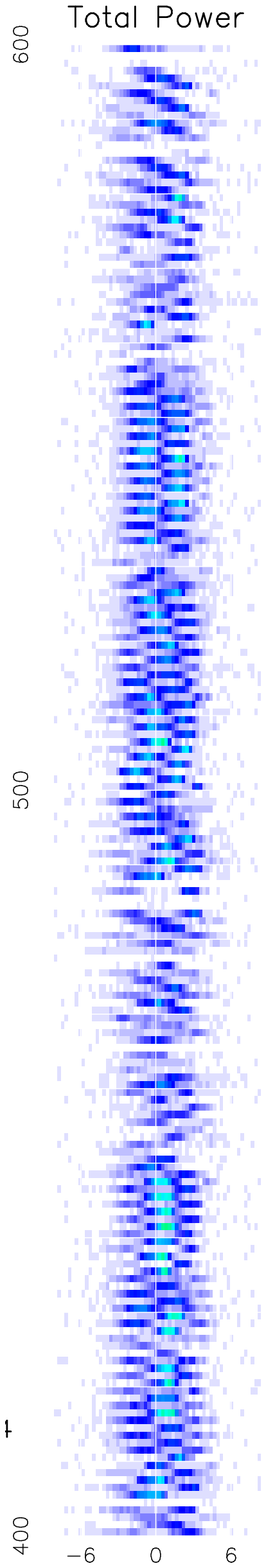,height=15.5cm}}}&
{\mbox{\psfig{file=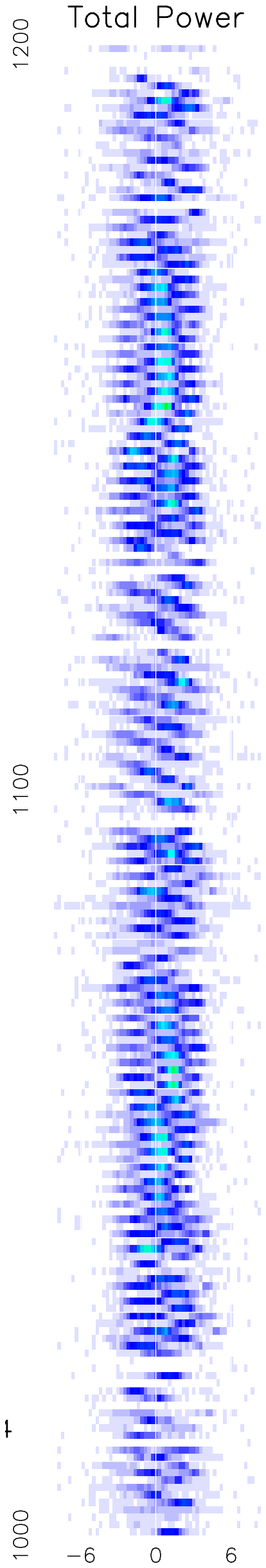,height=15.5cm}}}&
{\mbox{\psfig{file=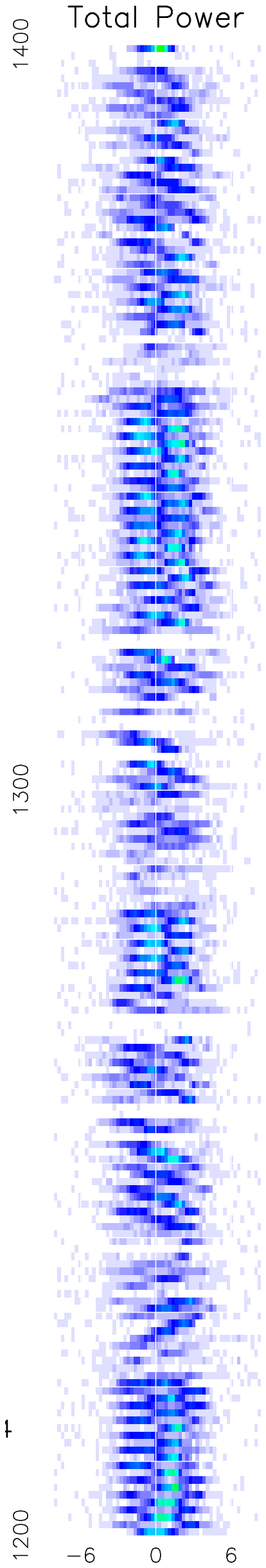,height=15.5cm}}}\\
\end{tabular}
\end{flushleft}
\caption{Subpulse sequences at 430 MHz. The colour scaling of Fig. 1 has been amended so that the changing subpulse phase is brought out more clearly. All pulse intensities are shown in shades of blue, with turquoise for the strongest. Note five examples of the unusual Q* mode, all following steady B-mode sequences. These are at pulses 450, 560, 1080, 1220 and 1362.}
\label{Fig116}
\end{figure}

 \subsection{Subpulse modulations}
To take our analysis one step deeper, we progress to the subpulse level. The subpulse modulations of each mode, succinctly summarised in the lrf and hrf spectra of Figs. 9 $\&$ 10, are revealed with great clarity in Fig 17, which shows three pulse sequences from the same 430-MHz observation. By suppressing the intensity scale so that all but the brightest pulses have the same colour, the two principal phase patterns of the modes are brought to the fore, and we can better study their interaction. Moreover, we can also see an additional mode, most unexpected and often barely discernible, which immediately follows five of the longer B-mode sequences. In this mode, which we call the Q* mode because of its `conjugate' nature, the driftbands have a reversed slope to those of the Q mode. The new mode, and its position in the sequence of modes, is the key to the model we develop here. 

We begin by considering the properties of the principal modes. Although the $P_3$ periodicities of both modes are subject to jitter and swing, the fact that one clusters around $2P_1$ and the other $3P_1$ suggests a remarkable harmonic relation, and this is supported by our mode simulation (Fig. 11). Other pulsars (with up to three drift modes) are also known to exhibit `harmonic' relationships between their discrete $P_3$ values ($\eg$ B0031--07, B1918+19, B1944+17, B2319+54, etc). However those pulsars have much longer $P_3$ periodicities (up to about $14P_1$) and their harmonic nature lies in the roughly equal $\it{ratios}$ of their $P_3$ values (around 3:2). B2303+30 not only has two $P_3$s with such a ratio --- it additionally has a harmonic relation $\emph{to the pulsar period itself}$. This presents us with what could be a major clue to their physical interpretation, if only it can be comprehended.

A further clue could be the curious fact that B2303+30 belongs to a small but possibly significant group of pulsars with a $P_3$ very close to $2P_1$. In general, older pulsars (measured by their spin-down rate) are found to have longer $P_3$ values (seen in Fig. 4 of Rankin 1986): a feature which is supported by more recent discoveries of subpulse drift in other pulsars, and it is surprising that there are now seven pulsars of all spindown ages which have $P_3$ close to the Nyquist value. The others are B0834+06, B0943+10, B1632+24, B1933+16, B2020+28, B2310+42. None of these pulsars are known at present to have a second Q-like fluctuation, but, with the exception of B0943+10 [whose  Q mode is chaotic (Rankin $\&$ Deshpande 2001)], little long-term single-pulse study has been undertaken of any of them.

Of course we cannot be sure that the $P_3$ values observed in B2303+30 are the true ones. We may be seeing aliases of a faster underlying drift, as has been suggested for the very different pulsar B0826--34 (Gupta \etal\  2004). However a number of arguments, of varying degrees of strength, can be assembled in favour of zero alias. Firstly, in the pulsar B0943+10, with which B2303+30 shares many properties, it was conclusively demonstrated (Deshpande $\&$ Rankin 2001) that its B-mode on-off emission was not generated by high-order aliasing [in fact it has the pattern of Fig. 18(a) and n=0 according to the alias ordering system here]. Secondly, if the observed $P_3$ values of both B and Q modes are the result of aliasing to the nth degree, their true $P_3$s would be $\frac{2}{2n+1}$ and $\frac{3}{3n+1}$ respectively, and their harmonic ratio to each other and to the rotation period would be weaker and more complex (and for high n, lost altogether). This would imply that the observed harmonic relations were coincidental. Furthermore, $\emph{differential}$ aliasing between the modes would result in complex transitions (van Leeuwen \etal\ 2002), yet at least some transitions (see Fig. 3) are observed to occur smoothly. Finally, if the modes were aliased to differing orders, then not only would simple harmonicity be lost, but the modes would have a differing $P_2$s (through a change in the number of rotating beams observed in a single turn), which is not observed.

Assuming the subpulse pattern has zero alias, we can determine the direction of movement of the subpulses in relation to the direction of the observer's sightline, and whether the underlying drift speeds up or slows down between modes. The fact that each of the modes has its characteristic observed $P_3$ enables us to constrain the possibilities, and Fig. 18 illustrates this point. In both schemes (a) and (b) the direction of the sightline is the same. In (a) (alias n=0) the underlying drift is counter to this direction and in (b) (alias n=--1) has the same sense. In the B-mode drift, assumed for simplicity to have an observed $P_3$ of exactly $2P_1$, the on-off system appears in both cases, but in (a) with the central subpulse repeating on the leading side as the drift progresses, and in (b) on the trailing side. If a secondary drift towards the trailing edge is observed (\eg pulses 60-90 in Fig. 1), this corresponds to an acceleration in case (a) (\ie \emph{true} $P_3 < 2P_1$) and a deceleration in case (b) (\emph{true} $P_3 > 2P_1$). Note that this argument does not actually depend on the beams forming the carousels shown in Fig. 18, nor on the number of such beams : it merely concerns the sequence in which the beams are presented to the observer. But if the beams do take the form of a carousel, then the line-of-sight traverse is interior in (a) and exterior in (b) to the angle between the magnetic and rotation axes, otherwise the net rotation of the carousel would exceed the rotation of the star.

In Section 5 we presented strong observational arguments for believing our sightline passes between the rotational axis and the magnetic axis (an inner traverse). These arguments were quite independent of the discussion in this Section and serve to confirm our preference for the subpulse system of Fig. 18(a). With this choice, we not only conclude that the elegant observed harmonic ratios correspond to true harmonics, but we find that our view of B2303+30's emission cone follows an inner traverse in essentially the same manner as B0943+10, and the subpulses of both pulsars are presented to us in the same sequence without alias.  

\begin{figure}
$$\vbox{
\psfig{file=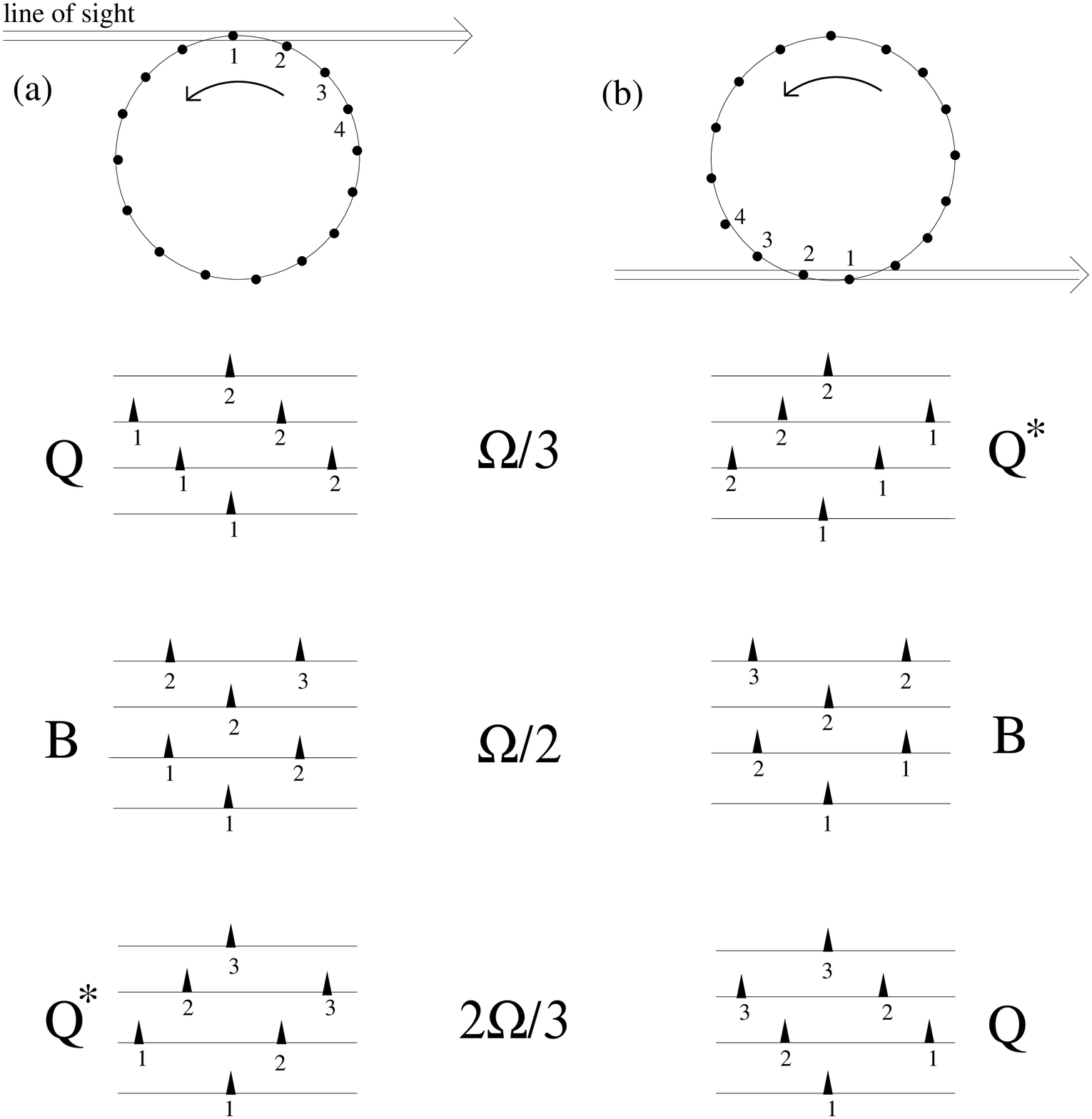,width=8truecm,angle=-0.}
}$$
\caption{The geometry of B- and Q-subpulse behaviour. The successive subbeams are represented as a carousel which rotates at speed $\Omega$, and the panels left and right show the resulting subpulse patterns for varying values of $\Omega$. In case (a) the carousel rotates counter to the sense of our sightline passage, and in (b) in the same sense, implying that (a) represents an observer traverse between the magnetic and rotation axes, and (b) outside both. The rarely-observed Q* mode has the same periodicity as mode Q but drifts in the opposite sense. Thus Q* is the aliased mode in (a) and Q in (b). In the text we argue that (a) is the more likely system for B2303+30 on observational grounds. The rotation rate $\Omega$ corresponds to the rate at which subbeam 2 would move to the position of subbeam 1 within one pulse period. }
\label{Fig18}
\end{figure}

\subsection{Mode transitions}
With the help of Fig. 18 we can examine how the B-mode system with $P_3\simeq2P_1$ migrates to a Q mode with $P_3=3P_1$ and the drift proceeding from trailing-to-leading edge [assuming no change in subpulse spacing]. The transition will be different in cases (a) and (b). The simplest ($n=1$) migration for (a) would be for the movement of the beams to slow to two-thirds from their B-mode speed (as shown). The subpulse arrangement would then imply that in the Q mode each subbeam was tracked across the drift band (the uppermost drift panel of Fig. 18(a)).  For the case (b) (\ie n negative) the simplest ($n=-1$) migration would be a small acceleration to 4/3 of its B driftrate, again leading to Q drift bands with $P_3=3P_1$, but aliased so that successive subpulses in a single band are successive subbeams (the lowest drift panel Fig. 18(b)). In either case, solutions with higher order alias are of course possible: the beams could accelerate from B to $(n-\frac{1}{3})$ times their original speed ($|n| > 1$), and thereby create an aliased Q drift with $P_3$ apparently $3P_1$. We argue against this above, but here we can be more specific: more highly aliased transitions would require the rotation to more than double and accelerate through a Nyquist boundary, resulting in an  apparently near-stationary drift. Careful inspection of the pulse sequences shows no convincing evidence of such drift patterns, and no low-frequency features  between 0.1 and 0.2 $c/P_{1}$ appear in the fluctuation spectra of Figs. 9 $\&$ 10. In short, we may reasonably assume that the transition from B to Q drift is the result of a change in the driftrate by either plus or minus $\frac{1}{3}$ of the B-mode driftrate.

In many cases the observed transitions from B drift to a regular Q drift and back are far from smooth. In Fig. 1 we can see that nulls or apparently disordered pulses often intervene. However there is a marked difference between B $\rightarrow$ Q  and Q $\rightarrow$ B transitions, as can be seen in the phase-enhanced sequences of Fig. 16. In the former case the B mode, shortly before the transition, is relatively steady and close to its $P_3\simeq2P_1$ state. Only in the B mode's last few pulses, if at all, does its aliased drift begin to strengthen towards the leading edge. This is then followed either by some disorder or by nulls or, in many cases, by drift $P_3\simeq3P_1$ $\emph{in the  opposite sense}$ to the standard Q drift, designated as Q* in Fig. 18. In the Q $\rightarrow$ B case, as the B mode recommences, there is often a strong drift from the trailing edge which mirrors the concluding behaviour of the B mode in its transition to Q, but here this effect is usually stronger and more marked, and continues with damped oscillating drift patterns well into the B mode.  Furthermore, no Q* drift is ever seen before a switch to the B mode. This clear distinction between the two kinds of transition gives the subpulse sequence a "time arrow". 

It is striking and unexpected that when steady B-mode drift begins the process which ultimately leads to the Q-mode drift, it first suddenly shifts its drift in a sense opposite to what we have identified as Q-mode drift.  This can be clearly seen after pulse 450 in the first sequence of Fig. 16, and at four other positions in these sequences. In terms of the carousel picture of Fig. 17(a) this requires that the subbeams accelerate from the configuration B away from Q towards the subpulse pattern Q*. In many cases this is exactly what is observed, although the acceleration is not always sufficient to achieve a clear Q* pattern. 

\begin{figure}
$$\mbox{
\psfig{file=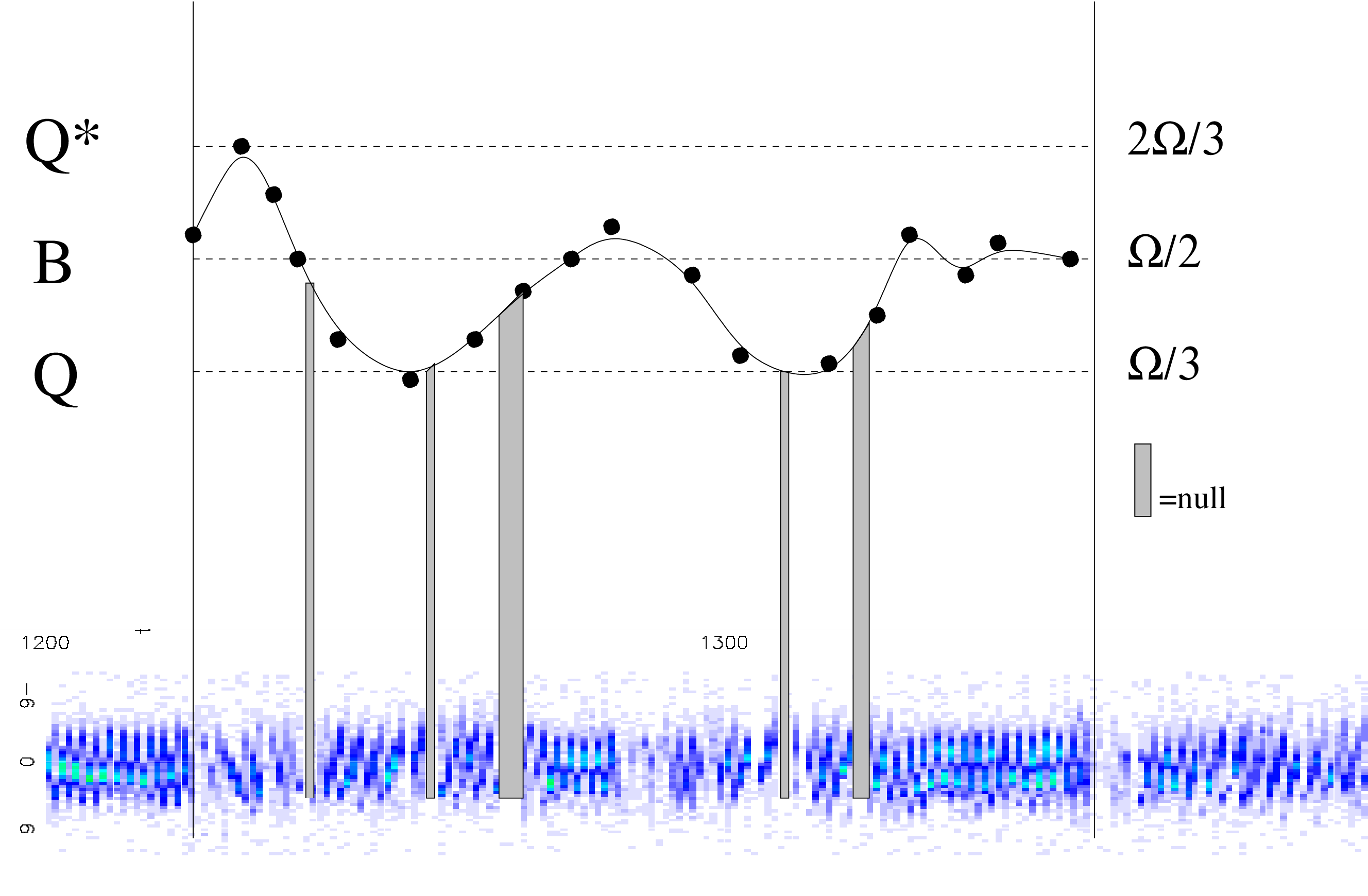,width=8truecm,angle=0.}
}$$
\caption{The varying driftrate following an `event' derived from the subpulse sequence in the right-hand panel of Fig. 17 and interpreted as subbeam rotation in terms of the scheme in Fig. 18(a). Note the sudden sharp increase in subpulse drift towards the Q* driftrate followed by a slow relaxation through successsive B mode and Q mode interludes, before returning to a sustained and weakly oscillating B mode. There is a suggestion that nulls may be associated with rapid driftrate changes.} 
\label{Fig. 19}
\end{figure}

\subsection{The B2303+30 system and `events'}
We are now in a position to interpret the constantly changing drift patterns in terms of varying subbeam rotation rates on the basis of the model in Fig. 18. The changes from a steady B mode sequence to Q are sometimes very sudden, and we designate these as `events', without prejudice to their underlying physical cause [this convenient nomenclature was introduced by Lyne $\&$ Ashworth (1983) in connection with the nulls of B0809+74, and also by Janssen $\&$ van Leeuwen (2004) for B0818--13]. Other changes are more gradual, and appear to be part of an ongoing process. As an illustration of our technique, in Fig. 19 we have analysed a pulse sequence extracted from the righthand column of Fig. 17 and interpreted by means of Fig. 18 as a single cycle. The main features are:

i) Starting in a steady B-mode sequence there is little hint of the impending mode change, just occasionally  a sudden shift in the aliased secondary drift towards the leading edge, before the B-mode intensity is lost. In other sequences this `warning' shift towards the end of B can be more gradual. 

ii) The Q mode commences as a single sudden event, causing an accelerated drift to the leading edge and a weakening in intensity,  sometimes apparently provoking a null sequence and/or an excursion to the Q*-mode driftrate. Here the effect is unusually powerful. According to the subpulse schemes in Fig. 18 this implies that the subpulses have either accelerated (a) or decelerated (b) $\emph{away}$ from the rotation rate corresponding to Q-mode drift. 

iii) Having achieved a peak  in the Q* sense, the driftrate begins to fall back, passing rapidly through the rate associated with B and moving towards a typical Q-mode rate. Throughout the entire cycle nulls are often noted at points where rapid change in drift can be inferred.

iv) Whether or not the Q driftrate is reached, the driftrate continues to oscillate with gradually reducing amplitudes around the mean B-mode value. 

v) Whenever Q-mode drift is achieved ($\eg$ pulses 1242-1265 and 1300-1320) the emission is relatively stable and bright, albeit sometimes interspersed with nulls.

vi) Finally the B mode is reestablished in the form of a weak damped wave in the driftrate around the B-mode value, which gradually peters out until the next event. These closing B-mode sequences are often approached by subpulse motion from the trailing edge (see pulse 1325), indicating that the driftrate has made a further oscillation towards Q* and not smoothly returned from the Q-drift sequence.

The sequence (i) to (vi) exemplifies the cyclic nature of this pulsar's behaviour. Although here at its most complex, it is possible to trace the swings in driftrate through 133 pulses. Other cycles are shorter and  may sometimes consist of just a single shallow oscillation, but all share the property of an event followed by a gradual relaxation to the B mode. The intensity of the events, measured by the amplitude of the initial perturbation in the driftrate towards (but only rarely achieving) the Q* pattern, can vary considerably. It is difficult to assess whether events are occurring at random in the pulse sequence (see Section 6.1), but the evidence of Fig. 18 and numerous other sequences suggest that the B mode is the underlying asymptotic steady state which, once achieved, triggers a fresh event in a never-ending feedback system.  

As a cycle progresses we see a sequence of short B and Q stretches and short nulls generated by the rapid oscillation in the subbeam-rotation rate. It is this which generates the coupled distributions of short B, Q and null sequences referred to in Section 6.1. Longer B sequences (peaking at 37$P_1$ in Fig. 4) represent relatively stable B conditions leading up to an event. Longer Q's (peak 31$P_1$) apply to the apparently confused sequences following an event, where, in the underlying cycle, the B driftrate is never sustained over sufficient time or intensity to be identified.

Thus we may conjecture that the emission of this pulsar is governed by a series of `events' : unspecified physical impulses which suddenly change the driftrate, and from which the emisson takes  many rotation periods to recover.  They appear to be triggered in the B mode by the very stability of its subpulse pattern, suggesting that the B mode is the pulsar's asymptotic equilibrium state upon which the events act. Following an event, the average relaxation time will be approximately the sum of the average durations of the B and Q modes in their longer manifestations (\ie the peaks on the broader mode length distributions in Fig. 4). This sum is around 80 pulses and generates the low frequency feature in Figs. 9 $\&$ 10.

\section{The magnetosphere of B2303+30} 
\subsection{Comparison with B0943+10}As pointed out earlier, B2303+30 is one of 7 pulsars which exhibit an on-off intensity variation with $P_3$ around $2P_1$.  The 7 have apparently little in common in their basic parameters of rotation period, inferred magnetic field strength and spin-down age so the effect may be a coincidence, but the wide range of the last of these parameters (from B0834+06 with a timing age of 3 Myr up to B1632+24 with 65 Myr) is interesting since it defies the empirical rule of most pulsars with subpulse drift [shown in fig. 4 of Rankin (1986)] that $P_3$ increases with age.

The pulsar from this group with parameters closest to B2303+30 is B0943+10, and this fact was initially the impulse for the present investigations. Many detailed investigations of B0943+10 have been published in recent years ($\eg$ Deshpande $\&$ Rankin 2001), its attraction being its precise $P_3$ modulation (close to $1.87P_1$) combined with a regular longer-term modulation, which have enabled these authors to establish uniquely that the pulsar fits a model of 20 emission columns circulating around the star's magnetic pole.

Both B0943+10 and B2303+30 have relatively long periods (1.1 s {\it vs.} 1.6 s --- and hence similar light-cylinder radii). They have comparable timing ages (5 Myr {\it vs.} 8.6 Myr) and they have near-identical inferred surface magnetic fields ($2\times{10^{12}}G$), so that their light cylinder magnetic fields only differ by a factor less than 3, and the magnetospheric electric field, which governs the driftrate (Ruderman $\&$ Sutherland 1975, Gil $\&$ Sendyk 2000, Wright 2003) will be of the same order in both pulsars. However there are fundamental differences. In the B mode of B2303+30 there is a clear `jitter' in the $P_{3}$ modulation, sometimes sudden and sometimes gradual but in any case within about 40 periods, which makes it impossible to uncover any circulatory modulations which the emission beams may possess and hence prevents us from determining the direction of the beams' rotation with respect to the observer. Curiously, our derived figure for the average duration of the B mode in B2303+30 ($\simeq{37P_1}$) is virtually identical to the circulation time given for the subbeam carousel in B0943+10 (37.35$P_1$). In any case the frequent "events" in B2303+30, which give rise to  fascinating interplay between B and Q modes, prevent the asymptotic B mode from being maintained for any great length of time.

The greatest contrast between the two pulsars lies in the behaviour of their respective Q modes: that of B0943+10 is undoubtedly highly chaotic with no periodic features  reliably detected so far (although, intriguingly, some remnant memory of the circulation timescale similar to the B mode does seem to persist).  In B2303+30 we have demonstrated beyond doubt that a clear and unambiguous drift with a periodicity of around $3P_1$ frequently occurs in its Q mode. Furthermore, nulls are found to be common in this mode (Fig. 11), and none have been detected in either mode of B0943+10. The Q mode of B2303+30 is marked by a considerable fall in intensity (Fig. 1). This is true of B0943+10's Q mode also, but there can be sudden powerful pulses and in general a greater `spikeyness' of emission. The reason for their differences might lie in their differing angles of inclination (in the model of Wright (2003) this is a critical factor in determining the characteristics of subpulse drift) or on the relative sizes of their emission cones. The angle for B0943+10 has been reliably fitted to about 15$\deg$ (Deshpande $\&$ Rankin 2001) on an outer cone. However, in Section 5, we argue, on the basis of the frequency dependence of its profile, that B2303+30 is at an angle of $26\deg$ and that we are seeing an inner cone. Nevertheless, our sightline traverses both pulsars on an inner passage between the magnetic pole and the rotation axis.

Relatively little is known about the relative occurrence of the two modes in B0943+10 since both modes typically last several hours, but it is likely to be more weighted towards B than the figure (54B : 46Q) obtained here for B2303+30 (Suleymanova \etal\ 2003). One interesting point of similarity is the observation in B0943+10 that the B mode anticipates its mode change to Q by slow changes in its driftrate (Suleymanova $\&$ Izvekova 1989). This can be seen as a more gradual version of the slight shift in the B drift pattern to the leading edge immediately before an event (see (i) in Section 6.4). On the whole one has the impression that B2303+30 is a speeded-up, more impatient, version of B0943+10: the modes switch much more frequently by a factor of at least 250 and the driftrates vary by a far greater factor. 

\subsection{Emission models}
Gil $\&$ Sendyk (2000) have applied their modified form of the Ruderman $\&$ Sutherland (1975) polar cap model to B2303+30. Their fit was based on two `sub-modes' of the B mode with $P_3$ just under $2P_1$ [\ie scheme (a) in Fig. 18]. They fitted a ring of 12 sparks on an outer cone with an interior traverse on an assumed axis inclination of $50\deg$. Our considerations in Section 5 put the lower figure of $26\deg$ on the inclination and suggest an inner cone, so their model may need refitting to check these changed parameters. The discovery here of a steady Q-mode drift substantially different from the B-mode drift will, in the context of a polar cap model, would require the surface electric field to make adjustments of up to $50\%$ in a very short time (less than one rotation period) and this is difficult to accommodate. However a revised version of the theory (Gil \etal\ 2003), wherein much of the electric field is screened close to the surface, may counter this objection.

A magnetosphere-wide feedback model can be constructed for the B-mode drift along the lines of the published fit to the B0943+10 drift (Wright 2003). This model regards the radio emission from close to the surface to be driven by particle infalls from the outer magnetosphere, particularly from a weakly pair-creating outer gap. The pattern of the subpulse drift in this model is highly dependent on the inclination angle, since it is this which determines the location of the outer gap, and predicts that stable slow drifting and slow long-term cycles are more likely in nearly aligned pulsars. More inclined pulsars will experience faster drift and more frequent moding due to the greater variations in the electric fields in the less stable magnetospheres of such pulsars (Rankin $\&$ Wright 2003). In B2303+30 the mode changes are relatively frequent, so an inclination in the $20\deg-30\deg$ range might be expected. In this model, a mode change might correspond to a shift in the location of the outer pair-creation site along the zero-charge surface, thereby changing the timescale of the feedback system. 

But what is the nature of the `events', which we have here identified as instigators of the mode changes? In B1237+25, a highly-inclined pulsar ($\alpha\simeq{53\deg}$), where we have the advantage of a sightline passage directly over the magnetic pole, the mode transition from an ordered to a disordered mode is always accompanied by activity in the core region of the profile close to the magnetic pole. These polar `eruptions' appear also to interfere with the normal mode on a quasi-periodic timescale of around 40$P_1$ (Hankins $\&$ Wright 1980, Srostlik $\&$ Rankin 2004). In B2303+30 our sightline cuts only the fringe of the emission cone, so perhaps the sudden `events' are due to unseen activity along the pulsar magnetic axis. On the other hand, the fitted driftrate curve in Fig.19 closely resembles those found in studies of non-linear damping of the potential in electrical systems [$\eg$ the van der Pol (1927) equation], where sudden changes in potential are just chaotic fluctuations in a completely deterministic system, and these require no `Deus ex machina'.

Neither model seems able to explain the harmonic relation between the observed $P_3=2P_1$ and $P_3=3P_1$ subpulse fluctuations. Noting that dipole geometry is built on the ratio 2:3  ($\eg$ the inner and outer radial limits of the `null' or zero net-charge line are in the ratio of $(\frac{2}{3})^{\frac{3}{2}}$), we can speculate that the harmonies might correspond to `eigensolutions' in the configuration of the magnetosphere at large.

\section{Conclusions}
We list the principle conclusions of this paper:

(1) B2303+30 has two modes of emission, B and Q, the former being more intense and more organised. The Q-mode profiles are slightly wider in shape than those of the B-mode at all frequencies, and the profiles of both modes are single-peaked and significantly asymmetric.  This asymmetry is also present in longitudinal cross-correlations and fluctuation spectra.

(2) Each of the modes exhibits a characteristic  subpulse behaviour. The B mode has an on-off pattern with a fluctuation frequency close to $2P_1$, sometimes steady and sometimes weakly modulated. The Q mode is more irregular, but often exhibits distinctive fluctuations with $P_3$ close to $3P_1$. There is no evidence of change in $P_2$, the subpulse spacing, from one mode to the other. Thus a harmonic relation exists between the modes' drift-rates and, quite remarkably, each mode is harmonically related to the pulsar rotation period.

(3) Nulls occur predominantly, possibly exclusively, within the Q-mode sequences, often close to the start or ending of the Q mode. This is the first pulsar in which nulls have been shown to be confined to a particular emission mode. There is evidence that nulls occur when the subpulse driftrate is undergoing rapid change. 

(4) All the observed subpulse features are closely knit into a single emission system. The system consists of a series of cycles, each begun by an `event', occurring when the B mode is relatively stable. This results in rapid driftrate changes, which we identify as the Q mode. Surprisingly, the event's initial effect, whether great or small, is generally counter to the variation required to bring about the Q-mode drift, and, when strong, achieves a subpulse pattern with $P_3\simeq3P_1$, as in the usual Q drift, but with the sense of drift reversed (this we call Q* drift).

(5) The driftrate then gradually relaxes back asymptotically to the steady B-mode driftrate, often in a roughly oscillatory manner over many pulses. The total relaxation period averages 80 pulses, and depends on the intensity of the event.  Swings in the driftrate lead to short alternating stretches of Q*, B and Q drift, interrupted by nulls which frequently occur when the underlying subbeam rotation is varying rapidly. Conversely, bright pulse sequences of either B or Q mode are corrrelated with slow variations in drift.

(6) The changing driftrates can be modelled as damped oscillations in the subbeam circulation rate. The circulation rate is found never to vary more than $\pm{\frac{1}{3}}$ from that of the B-mode. In the simplest case (our preferred solution) the subbeam rotation is counter to the sense of the observer's sightline in an $\emph{internal}$ traverse [Fig. 18(a)]. This implies that the Q* drift is aliased, but B and Q are not : the change from B- to Q-mode drift slows the subbeam rotation to two thirds of its B rate, whereas directly after `events' the circulation rate speeds up (by one third in the case of Q*). Then the $\emph{true}$ B-mode $P_3$ is just below $2P_1$, as in the case of B0943+10, and the  mean $\emph{true}$ $P_3$ for Q is exactly $3P_1$. 

(7) The geometry of the circulation model is independently supported by the evidence from the frequency dependence of the pulsar's profile. This suggests that B2303+30 is inclined at an angle of $26\deg$ and that our sightline makes an interior traverse of an inner cone.   

(8) The B mode can be seen as the pulsar's asymptotic steady state, which nevertheless cannot be sustained for long without triggering an event. There is evidence (Fig. 16) that the occurrence and intensity of the events are not randomly distributed, and may themselves have a quasi-cyclic behaviour.

\noindent {\bf Acknowledgments:}
We thank Svetlana Suleymanova for discussions and Avinash Deshpande for assistance with 
aspects of the observations and analyses. One of us (SLR) wishes to acknowledge a HeliX/EPSCOR Summer Research Fellowship in partial support of this work, and another (GAEW) is grateful to the Astronomy Centre of the University of Sussex for the award of a Visiting Research Fellowship, and thanks the University of Vermont for a Visiting Scholarship during much of this work. Portions of this work were carried out with support from US National Science Foundation Grant AST 99-87654.  Arecibo Observatory is operated by Cornell University under contract to the US NSF.


\begin{thebibliography}{999}

\bibitem[]{} Backer, D. C. 1973, \apj, 182, 245.
\bibitem[]{} Deshpande, A. \& Rankin, J.M. 1999, \apj, 524, 1008
\bibitem[]{} Deshpande, A. \& Rankin, J.M. 2001, \mnras, 322, 438.
\bibitem[]{} Filippenko, A.V., Radhakrishnan, V. 1982, \apj, 263, 828
\bibitem[]{} Gil, J.A., \& Sendyk, M.  2000, \apj, 541, 351.
\bibitem[]{} Gil, J.A., Snakowski, J.K., \& Stinebring, D.R. 1992 \aap, 242, 119.
\bibitem[]{} Gil, J.A., Hankins, T.H., Nowakowski, L.  1992, in  IAU Colloq.182, The Magnetospheric Structure and Emission Mechanisms of Radio Pulsars. Pedagogical University Press, eds Hankins, T.H., 	Rankin,J.M.,Gil,J., Zielona Gora, 278.
\bibitem[]{} Gil, J.A., Lyne, A.G., Rankin, J.M., Snakowski, J.K., \& Stinebring, D.R. 1992 \aap, 255, 181.
\bibitem[]{} Gil, J.A., Melikidze, G.I., \& Geppert, U.  2003 \aap, 407, 315.
\bibitem[]{} Gupta, Y., Gil, J., Kijak, J., Sendyk, M. 2004, astro-ph/0404216
\bibitem[]{} Hankins, T.H., \& Rankin 2004, preprint
\bibitem[]{} Hankins, T.H., \& Wolszan, A. 1987, \apj, 318, 410.
\bibitem[]{} Hankins, T.H., \& Wright, G.A.E., 1980, \nat, 288, 681.
\bibitem[]{} Harding, A., Muslimov, A. 2002, \apj, 568,864
\bibitem[]{} Hibschmann, J., Arons, J. 2001, \apj, 560, 871
\bibitem[]{} Janssen, G., \& van Leeuwen, J. 2004, astro-ph/0406486.
\bibitem[]{} Jones, P.B. 1982,\mnras, 200, 1081
\bibitem[]{} Lang, K. R. 1969, \apj, 158, 175.
\bibitem[]{} Lyne, A.G., \& Ashworth, M. 1983, \mnras, 204,519.
\bibitem[]{} Lyne, A.G., \& Manchester, R.N. 1988, \mnras, 234, 477
\bibitem[]{} Malofeev, V, M., Izvekova, V. A., \& Shitov, Yu. P. 1989, AZh, 66, 345.
\bibitem[]{} Oster L., Hilton D.A., \& Sieber W.  1977, \aap, 57, 323.
\bibitem[]{} Ramachandran, R., Rankin, J.M., Stappers, B.W., Kouwenhoven,ÊM.L.A., \& vanÊLeeuwen,ÊA.G.J. 2002, \aap, 381, 993
\bibitem[]{} Ramachandran, R., Backer, D.C., Rankin, J.M., Weisberg, J.M., \& Devine, K.E.  2004, \apj, 606, 1167.
\bibitem[]{} Rankin, J.M. 1983, \apj, 274, 333 .
\bibitem[]{} Rankin, J.M. 1986, \apj, 301, 901.
\bibitem[]{} Rankin, J.M. 1990 (RI), \apj, 352, 247.
\bibitem[]{} Rankin, J.M. 1993a (RII), \apj, 405, 285.
\bibitem[]{} Rankin, J.M. 1993b, \apj.Suppl., 85, 145.
\bibitem[]{} Rankin, J.M., Campbell, D.B., \& Backer, D.C. 1974, \apj, 188, 609.
\bibitem[]{} Rankin, J.M., \& Ramachandran 2003, \apj, 590, 411
\bibitem[]{} Rankin, J.M., \& Wright, G. A. E. 2003, \aap, Rev, 12, 43.
\bibitem[]{} Ruderman, M.A., \& Sutherland, P.G. 1975 (RS), \apj, 196, 51.
\bibitem[]{} Sieber, W., \& Oster, L. 1975, \aap, 38, 325.
\bibitem[]{} Srostlik, Z. \& Rankin, J.M., 2004, in preparation.
\bibitem[]{} Suleymanova, S.A. \& Izvekova, 1989, \mnras,
\bibitem[]{} Suleymanova, S.A., Izvekova, V.A., Rankin, J.M., \& Rathnasree, N. 1998, J. Astr. \& Ap., 18, 1.
\bibitem[]{} Suleymanova, S.A. \& Rankin, J.M. 2004, in preparation
\bibitem[]{} Suleymanova, S.A. 2004, private communication.
\bibitem[]{} Takens, F. 1981, in "Dynamical Systems and Turbulence", eds. Rand, D.A. \& Young, L.-S., New York: Springer-Verlag. Lecture Notes in Math, 898, 366.
\bibitem[]{} Taylor, J.H., \& Huguenin, G.R. 1971, \apj, 167, 273
\bibitem[]{} van der Pol, B. 1927, Phil Mag, 3, 65
\bibitem[]{} van Leeuwen, A.G.J., Kouwehoven, M.L.A., Ramachandran, R., Rankin, J.M., Stappers, B.W. 2002, \aap, 387, 169
\bibitem[]{} Wolszczan A. 1980, \aap, 86, 7.
\bibitem[]{} Wright,G.A.E. 2003, \mnras, 344, 1041.
\bibitem[]{} Wright, G.A.E., \& Fowler, L.A. 1981a, \aap, 101, 356.
\bibitem[]{} Wright, G.A.E., \& Fowler, L.A. 1981b, in IAU Symp. 95, Pulsars, eds W.Sieber $\&$ R.Wielebinski, Dordrecht:Reidel, 221

\end{thebibliography}
\end{document}